\documentclass{PoS}

\usepackage{amsmath}
\usepackage[utf8]{inputenc}

\newcommand{\Tr}{\mathrm{Tr}\,}

\title{Family symmetries and CP}

\ShortTitle{Family symmetries and CP}

\author{\speaker{I. de Medeiros Varzielas}%
         \\
        University of Southampton\\
        E-mail: \email{ivo.de@soton.ac.uk}}

\abstract{CP-odd invariants, independent of basis and valid for any choice of CP transformation are a powerful tool in the study of CP. They are particularly convenient to study the CP properties of models with family symmetries.
After interpreting the consequences of adding specific CP symmetries to a Lagrangian invariant under $\Delta(27)$, I use the invariant approach to systematically study Yukawa-like Lagrangians with an increasing field content in terms of $\Delta(27)$ representations.
Included in the Lagrangians studied are models featuring explicit CP violation with calculable phases (referred to as explicit geometrical CP violation) and models that automatically conserve CP, despite having all the $\Delta(27)$ representations.}

\FullConference{18th International Conference From the Planck Scale to the Electroweak Scale \\
		 25-29 May 2015\\
		 Ioannina, Greece }

\begin{document}

\section{Introduction}

This contribution to the proceedings of Planck 2015 follows closely the layout of the seminar I presented in the conference. I include here an expanded discussion of situations with multiple $\Delta(27)$ singlets and triplets, studied recently in \cite{Branco:2015hea, Varzielas:2015fxa, Branco:2015gna}.

\subsection{Why study CP?}

Flavour is an unsolved problem in the Standard Model (SM) and the same can be said of CP phenomena, which are currently not well understood. When combined these constitute the flavour and CP problems of the SM but also of extensions like Supersymmetry.

The Baryon Asymmetry of the Universe can not be quantitatively accounted for in the SM, and the experimental bound on CP violation in the strong sector is extremely small. In the SM there is CP violation only in association with the Yukawa couplings of the quark sector, although experimental verification of CP violation in the lepton sector may soon be achieved by the increasingly precise neutrino oscillation experiments.

It is very timely to consider what are the most promising solutions to these kind of problems and possibly make predictions of what would be the observed phases in the PMNS leptonic mixing matrix. A recent ambitious example is the $A_4 \times SU(5) \times$ CP model studied in \cite{Bjorkeroth:2015ora, Bjorkeroth:2015tsa}, which simultaneously solves the strong CP problem, predicts all the CP phases of the PMNS and, through leptogenesis, links this prediction with the  Baryon Asymmetry of the Universe.

Given that there are good reasons to study CP, I now consider how one may do so.

\subsection{The invariant approach}

The invariant approach (IA) to CP is not new \cite{Bernabeu:1986fc}.
It starts by splitting the Lagrangian into $\mathcal{L}_{CP}$, that automatically conserves CP (e.g. kinetic terms, gauge interactions) and $\mathcal{L}_{rem.}$, the remaining part:
\begin{equation}
\mathcal{L}=\mathcal{L}_{CP}+\mathcal{L}_{rem.} \,.
\end{equation}
Subsequently:
\begin{itemize}
\item Impose the most general CP transformations (that leave $\mathcal{L}_{CP}$ invariant).
\item Apply them and see if it restricts $\mathcal{L}_{rem.}$.
\end{itemize}
The possibility of (explicit) CP violation only exists in the Lagrangian if the most general CP transformations constrain the Lagrangian (i.e. restrict $\mathcal{L}_{rem.}$).

The IA is powerful because:
\begin{itemize}
\item Gets results just from the Lagrangian.
\item Independent of basis.
\item Shows relevant quantities for physical processes.
\end{itemize}

A review of the IA for SM leptons is present in \cite{Branco:2015hea, Varzielas:2015fxa}.

\section{Invariant approach and family symmetries}

As shown by its innovative application to cases with family symmetries \cite{Branco:2015hea}, the IA proves to be particularly useful because the CP-odd invariants (CPIs) can be constructed directly from the Lagrangian, without knowledge of the family symmetry, and then used together with the specific structures enforced by the family symmetry on e.g. the Yukawa couplings. This becomes clearer when discussing specific examples, focused on trilinear terms which I refer to as Yukawa-like couplings, as most cases I consider here are meant as fermion-fermion-scalar terms.

\subsection{Discrete groups}

An interesting example of the use of CPIs with discrete groups arises from applying a relevant SM lepton sector CPI, $I_1$, constructed similarly to the quark sector CPI in \cite{Bernabeu:1986fc}.
Defining the Hermitian combinations $H_\nu \equiv m_\nu m_\nu^\dagger$ and $H_l \equiv m_l m_l^\dagger$ \cite{Branco:2015hea}:
\begin{equation}
I_1 \equiv \Tr \left[H_\nu , H_l \right]^3\,.
\label{hhcube}
\end{equation}
It turns out that $I_1$ is useful to analyse a Lagrangian with $A_4$ family symmetry determining the mass structures to be \cite{Branco:2015hea}:
\begin{equation}
m_{\nu}= \alpha \begin{pmatrix} 2&-1&-1\\-1&2&-1\\-1&-1&2\end{pmatrix}
+\beta  \begin{pmatrix}1&0&0\\0&0&1\\0&1&0 \end{pmatrix} \,, \quad \beta
+\gamma \begin{pmatrix}0&0&1\\0&1&0\\1&0&0 \end{pmatrix}
+\delta \begin{pmatrix} 0&1&0\\1&0&0\\0&0&1\end{pmatrix},
\label{mR}
\end{equation} 
in a basis where $m_l$ is diagonal.
The structures $\beta$, $\gamma$, $\delta$ each correspond to contractions to a different singlet of $A_4$ ($\beta$ corresponds to the trivial singlet).
With $H_l$ diagonal, $I_1$ is
\begin{equation}
I_1
=6i(m_{\mu}^2-m_e^2)(m_{\tau}^2-m_e^2)(m_{\tau}^2-m_{\mu}^2) \mathrm{Im} (H^{21}_{\nu}H^{13}_{\nu}H^{32}_{\nu}).
\end{equation}
CP conservation requires $I_1=0$ and since there are no mass degeneracies the relevant quantity is:
\begin{equation}
\mathrm{Im} (H^{21}_{\nu}H^{13}_{\nu}H^{32}_{\nu})=-\mathrm{Im} (\beta \delta^* + \gamma \beta^* + \delta \gamma^*) \mathrm{Re} (R)
\label{I1A4}
\end{equation}
where $R$ is a rather complicated expression,
\begin{eqnarray}
R&=&27|\alpha|^4-6|\alpha|^2|\beta + \gamma + \delta|^2+|\gamma \delta|^2+|\delta \beta|^2+|\beta \gamma |^2
\nonumber \\
&+& 4|\beta|^2 (\gamma \delta^*)+4|\gamma|^2 (\delta \beta^*)+4|\delta|^2 (\beta \gamma^*)\nonumber \\
&+&-6\alpha^{*2}(\beta^2+\gamma^2+\delta^2-\beta \gamma - \delta \beta - \gamma \delta)\nonumber \\
&+& 2\beta^{*2}(\gamma^2+\delta^2+\gamma \delta) 
+ 2\gamma^{*2}(\delta^2+\delta \beta) + 2\delta^{*2}\beta \gamma  \nonumber .
\end{eqnarray}
The conclusion is a result known previously in the literature, that this type of $A_4$ model automatically conserves CP, in the presence of only 1 singlet (this corresponds effectively to having 2 out of $\beta$, $\gamma$, $\delta$ equal to zero and therefore $I_1 = 0$).
This brief $A_4$ example also serves to show that the IA is useful beyond the $\Delta(27)$ cases which I focus on here.

\subsection{$\Delta(27)$}

In the following sections some knowledge of $\Delta(27)$ is useful.
I define $\omega \equiv e^{i 2 \pi/3}$, $c$ (for cyclic) and $d$ (for diagonal) as the relevant generators ($\omega^3=1$, $c^3=d^3=1$). The irreducible representations are 1 or 3 dimensional - singlets and triplets. Generators act on singlets by multiplying with a phase: $c 1_{ij}=\omega^i 1_{ij}$ and $d 1_{ij}=\omega^j 1_{ij}$, where $i, j = 0, 1, 2$ for a total of 9 singlets. In a suitable basis the generators act on a $3_{01}$ triplet $A=(a_1,a_2,a_3)_{01}$ or a $3_{02}$ triplet $\bar{B}=(\bar{b}^1,\bar{b}^2,\bar{b}^3)_{02}$ as:
\begin{equation}
c_{3_{0j}}=
\begin{pmatrix}
	0 & 1 & 0 \\
	0 & 0 & 1 \\
	1 & 0 & 0
\end{pmatrix}
 \,, \quad
c_{3_{01}}
\begin{pmatrix}
	a_1 \\
	a_2 \\
	a_3
\end{pmatrix}
=
\begin{pmatrix}
	a_2 \\
	a_3 \\
	a_1
\end{pmatrix} \,,
\end{equation}

\begin{equation}
d_{3_{01}}=
\begin{pmatrix}
	1 & 0 & 0 \\
	0 & \omega & 0 \\
	0 & 0 & \omega^2
\end{pmatrix} \,, \quad
d_{3_{02}}=
\begin{pmatrix}
	1 & 0 & 0 \\
	0 & \omega^2 & 0 \\
	0 & 0 & \omega
\end{pmatrix} \,.
\end{equation}
The nomenclature for the generators represents their action on triplets. $d$ distinguishes $3_{01}$ and $3_{02}$ according to their subscripts, which are the powers of $\omega$ on the first two diagonal entries of the respective matrix. I refer to $3_{01}$ as the triplet representation and to $3_{02}$ as the anti-triplet representation. $c$ cyclically permutes the components equally for triplet and anti-triplet.

Singlet product leads to a singlet transforming with the sum of the indices (modulo 3): $1_{ij} \times 1_{kl}$ transforms as $1_{(i+k) (j+l)}$. The product of triplet with anti-triplet is the sum of all nine singlets, including the trivial singlet
\begin{align}
[A \bar{B}]_{00} \equiv (a_1 \bar{b}^1 + a_2 \bar{b}^2 + a_3 \bar{b}^3)_{00} \,,
\label{AB00}
\end{align}
which is the $SU(3)$ invariant contraction, and the 8 non-trivial singlets
\begin{align}
[A \bar{B}]_{01} &\equiv (a_1 \bar{b}^3 + a_2 \bar{b}^1 + a_3 \bar{b}^2)_{01} \label{AB01} \,, \\
[A \bar{B}]_{02} &\equiv (a_1 \bar{b}^2 + a_2 \bar{b}^3 + a_3 \bar{b}^1)_{02} \label{AB02} \,, \\
[A \bar{B}]_{10} &\equiv (a_1 \bar{b}^1 + \omega^2 a_2 \bar{b}^2 + \omega a_3 \bar{b}^3)_{10} \label{AB10} \,, \\
[A \bar{B}]_{11} &\equiv (\omega a_1 \bar{b}^3 + a_2 \bar{b}^1 + \omega^2 a_3 \bar{b}^2)_{11} \label{AB11} \,, \\
[A \bar{B}]_{12} &\equiv (\omega^2 a_1 \bar{b}^2 + \omega a_2 \bar{b}^3 + a_3 \bar{b}^1)_{12} \label{AB12} \,, \\
[A \bar{B}]_{20} &\equiv (a_1 \bar{b}^1 + \omega a_2 \bar{b}^2 + \omega^2 a_3 \bar{b}^3)_{20} \label{AB20} \,, \\
[A \bar{B}]_{21} &\equiv (\omega^2 a_1 \bar{b}^3 + a_2 \bar{b}^1 + \omega a_3 \bar{b}^2)_{21} \label{AB21} \,, \\
[A \bar{B}]_{22} &\equiv (\omega a_1 \bar{b}^2 + \omega^2 a_2 \bar{b}^3 + a_3 \bar{b}^1)_{22} \label{AB22} \,.
\end{align}

\subsubsection{$\Delta(27)$ and adding CP} 

I will now study the CP properties of a specific $\Delta(27)$ invariant Lagrangian with a triplet $A$, anti-triplet $\bar{B}$, and singlets $C$, $D$ (respectively $3_{01}$, $3_{02}$, $1_{10}$, $1_{01}$):
\begin{equation}
\mathcal{L}_{CD}=y_c (A \bar B)_{20} C_{10} + y_d (A \bar B)_{02} D_{01} + H.c. \,.
\end{equation}
Now I add a specific CP transformation, like the trivial CP transformation $CP_1$ acting $A$, $\bar{B}$, $C$ and $D$ as:
\begin{eqnarray}
CP_1 A &=& A^*  = (a^{1*},a^{2*},a^{3*})_{02} \,, \\
CP_1 \bar{B} &=& \bar{B}^*= (\bar{b}_1^*,\bar{b}_2^*,\bar{b}_3^*)_{01} \,, \\
CP_1 C_{10} &=& C_{20}^* \,,\\
CP_1 D_{01} &=& D_{02}^* \,.
\end{eqnarray}
Note that $A^*$, $\bar{B}^*$, $C^*$, $D^*$ are respectively $3_{02}$, $3_{01}$, $1_{20}$, $1_{02}$ (reflected by the indices and labels).

Imposing invariance under $CP_1$ on $\mathcal{L}_{CD}$,
the $y_c$ term which $CP_1$ transforms to:
\begin{equation}
\to y_c (a^{1*} \bar{b}_1^* + \omega a^{2*} \bar{b}_2^* + \omega^2 a^{3*} \bar{b}_3^*)_{20} C_{20}^* \,,
\label{CP1c}
\end{equation}
should become the $H.c.$, with $y_c^*$:
\begin{equation}
y_c^* (a^{1*} \bar{b}_1^* + \omega^2 a^{2*} \bar{b}_2^* + \omega a^{3*} \bar{b}_3^*)_{10} C_{20}^* \,.
\end{equation}
Beyond the coefficient being conjugated, the expressions are different (noted by their labels).
Instead, $CP_1$ transforms the $y_d$ into:
\begin{equation}
\to y_d (a^{1*} \bar{b}_2^* + a^{2*} \bar{b}_3^* + a^{3*} \bar{b}_1^*)_{01} D_{02}^* \,,
\end{equation}
which compares to its $H.c.$ with $y_d^*$:
\begin{equation}
y_d^* (a^{1*} \bar{b}_2^* + a^{2*} \bar{b}_3^* + a^{3*} \bar{b}_1^*)_{01} D_{02}^* \,.
\end{equation}
Apart from conjugating $y_d$ the expressions are the same.
This reveals that the $CP_1$ transformed expression is not invariant under $\Delta(27)$ for arbitrary $y_c$ (adding the subscripts will not make a trivial singlet). $\mathcal{L}_{CD}$ is only invariant under both $\Delta(27)$ and $CP_1$ if $y_c=0$ (and $y_d$ to be real) or conversely, keeping $y_c \neq 0$ explicitly violates either $\Delta(27)$ or $CP_1$.

Although imposing a specific CP transformation can force coefficients to vanish, this does not imply CP violation occurs if those coefficients do not vanish. $\mathcal{L}_{CD}$ with arbitrary $y_c$ and $y_d$ is actually CP conserving.
More considerations on adding CP to family symmetries and $\Delta(27)$ in particular can be found in \cite{Varzielas:2015fxa}, where changes of basis are considered.

\subsubsection{$\Delta(27)$ just singlets}

To illustrate how the IA proceeds, I start with Yukawa-like terms without $\Delta(27)$ triplets. I name singlets under $\Delta(27)$ $h_{ij}$, the label means it is a $1_{ij}$.
With $h_{00}$, $h_{01}$, $h_{10}$, the Yukawa-like terms are \cite{Branco:2015gna}:
\begin{align}
\mathcal{L}_{III} =  & z_{00} h_{00} h_{00} h_{00} + z_{01} h_{01} h_{01} h_{01} + z_{10} h_{10} h_{10} h_{10} \notag \\
+ & y_{00} h_{00} h_{00} h_{00}^\dagger + y_{01} h_{00} h_{01} h_{01}^\dagger + y_{10} h_{00} h_{01} h_{01}^\dagger + H.c. \,.
\label{Ls}
\end{align}
The next step is to consider the most general CP transformation for each field, each singlet transforms with its own phase $p_{ij}$
\begin{align}
h_{ij} \rightarrow e^{i p_{ij}} h_{ij}^*  \,.
\label{hij}
\end{align}
For $\mathcal{L}_{III}$ to remain invariant under these transformations leads to a set of necessary and sufficient conditions for CP conservation
\begin{align}
z_{00} e^{i 3 p_{00}}= z_{00}^* \,, \quad z_{01} e^{i 3 p_{01}}= z_{01}^* \,,\quad z_{10} e^{i 3 p_{10}}= z_{10}^* \,, \\
y_{00} e^{i p_{00}}= y_{00}^* \,,\quad y_{01} e^{i p_{00}}= y_{01}^* \,,\quad y_{10} e^{i p_{00}}= y_{10}^* \,.
\label{ys}
\end{align}
By combining conditions that cancel dependence on the CP transformations one obtains CPIs. A CPI with $y_{01}$ and $y_{10}$ requires canceling the dependence on $p_{00}$, as in $\mathrm{Im} [ y_{01} y_{10}^*]$
\begin{equation}
y_{01} y_{10}^\dagger = (y_{01} y_{10}^\dagger)^* \to
\mathrm{Im} [ y_{01} y_{10}^*] = 0 \,,
\label{0110_rel}
\end{equation}
where $y_{ij}$ are complex numbers ($y_{ij}^\dagger = y_{ij}^*$).
The CPI vanishing is necessary (but often not sufficient) for CP conservation, and in this case constrains the relative phase between the couplings.

I generalise the field content to include all 9 $\Delta(27)$ singlets $h_{ij}$. Imposing a $Z_3$ symmetry where each $h_{ij}$ transforms equally can reduce the allowed terms. There are 9 Yukawa-like terms like $z_{00} h_{00} h_{00} h_{00}$ (one for each singlet) but I focus on the mixed terms like $y_{1} h_{00} h_{01} h_{02}$, of which there are 12 combinations \cite{Branco:2015gna}:
\begin{align}
\mathcal{L}_{IX} =  y_{1} h_{00} h_{01} h_{02} + y_{2} h_{00} h_{10} h_{20} + y_{3} h_{00} h_{11} h_{22} + y_{4} h_{00} h_{12} h_{21}+& \notag \\
 y_{5} h_{01} h_{10} h_{22} + y_{6} h_{01} h_{11} h_{21} + y_{7} h_{01} h_{12} h_{20} +& \notag \\
 y_{8} h_{02} h_{10} h_{21} + y_{9} h_{02} h_{11} h_{20} + y_{10} h_{02} h_{12} h_{22} +& \notag \\
 y_{11} h_{10} h_{11} h_{12} + y_{12} h_{20} h_{21} h_{22} +& H.c. \,. \label{Ls9}
\end{align}
The CP conservation condition for each coupling depends on the 3 phases of the respective singlets:
\begin{align}
y_{1} e^{i (p_{00}+p_{01}+p_{02})} = y_{1}^* \,, \quad y_{2} e^{i (p_{00}+p_{10}+p_{20})} = y_{2}^* \,, \quad y_{6} e^{i (p_{01}+p_{11}+p_{21})} = y_{6}^* \,, \\
y_{10} e^{i (p_{02}+p_{12}+p_{22})}= y_{10}^* \,, \quad y_{11} e^{i (p_{10}+p_{11}+p_{12})}= y_{11}^* \,, \quad y_{12} e^{i (p_{20}+p_{21}+p_{22})}= y_{12}^* \,.
\label{ys9}
\end{align}
It is possible to combine several of the mixed couplings to form a CPI. An example is
\begin{equation}
\mathrm{Im} [y_1 y_2^* y_6^* y_{10}^* y_{11} y_{12}] \,,
\end{equation}
so this particular combination of couplings has to be real for CP to be conserved. Other combinations can be built from the couplings in $\mathcal{L}_{IX}$.

\subsubsection{$\Delta(27)$ pair of triplets}

The next case study for the IA are Yukawa-like terms with 2 $\Delta(27)$ triplets (the case with 1 $\Delta(27)$ triplet can be found in \cite{Varzielas:2015fxa, Branco:2015gna}). An interesting Lagrangian is similar to $\mathcal{L}_{CD}$: 
\begin{equation}
\mathcal{L}_{3s} = y_{00} (L \nu^c)_{00} h_{00} + y_{01} (L \nu^c)_{02} h_{01} + y_{10} (L \nu^c)_{20} h_{10} + H.c. \,,
\end{equation}
where there are now 3 singlets, and $L$, $\nu^c$ are the triplet and anti-triplet respectively.
The general CP transformations are associated with unitary transformations:
\begin{align}
&h_{00} \rightarrow e^{i p_{00}} h_{00}^* ;\quad
h_{01} \rightarrow e^{i p_{01}} h_{01}^*; \quad
h_{10} \rightarrow e^{i p_{10}} h_{10}^*; \quad
\\ 
&L \rightarrow U_L^T L^* ;\quad
\nu^c \rightarrow U_\nu \nu^{c *}\,.
\end{align}
Identifying the Yukawa matrices associated with each $h_{ij}$ as $Y_{ij}$, CP invariance requires
\begin{align}
U_L Y_{ij} U_\nu e^{i p_{ij}}= Y_{ij}^* \,.
\label{Y3s}
\end{align}

The relevant CPI is \cite{Branco:2015hea}:
\begin{equation}
I_{3s} \equiv \mathrm{Im} \Tr (Y_{00} Y_{01}^\dagger Y_{10} Y_{00}^\dagger Y_{01} Y_{10}^\dagger) \,.
\label{I3s}
\end{equation}
Note that this invariant applies to the Lagrangian even in the absence of $\Delta(27)$. 
However, $\Delta(27)$ invariance imposes additionally $Y_{00}=y_{00} \mathrm{I}$ (proportional to the identity matrix) and
\begin{equation}
Y_{01}=y_{01}
\begin{pmatrix}
0 & 1 & 0\\
0 & 0 & 1\\
1 & 0 & 0
\end{pmatrix}  ; \quad
Y_{10}=y_{10}
\begin{pmatrix}
1 & 0 & 0\\
0 & \omega & 0\\
0 & 0 & \omega^2
\end{pmatrix} \,.
\end{equation}

If one calculates the CPI for the $\Delta(27)$ invariant Lagrangian one obtains:
\begin{equation}
I_{3s}=\mathrm{Im} (3 \omega^2 |y_{00}|^2  |y_{01}|^2 |y_{10}|^2) \,,
\end{equation}
where the only phase present is $\omega^2$. The IA reveals a case of geometrical CP violation, i.e. where CP is violated but the arbitrary phases of the couplings (in this case the $y_{ij}$) are irrelevant.

Note that this type of invariant can only be built with 3 or more Yukawa matrices, which is a hint that cases with 2 singlets automatically conserve CP - as is the case for $\mathcal{L}_{CD}$ and is shown in \cite{Varzielas:2015fxa}.
In fact there is explicit geometrical CP violation for Lagrangians of type $\mathcal{L}_{3s}$ with almost any combination of 3 $\Delta(27)$ singlets \cite{Varzielas:2015fxa} - the exceptions are when choosing one of the 12 combinations of 3 singlets that make up an invariant term in $\mathcal{L}_{IX}$, in such cases the Lagrangian conserves CP automatically. One such example is:
\begin{equation}
\mathcal{L}_{3s1} = y_{00} (L \nu^c)_{00} h_{00} + y_{01} (L \nu^c)_{02} h_{01} + y_{02} (L \nu^c)_{01} h_{02} + H.c. \,,
\end{equation}
associating with each $h_{ij}$ as $Y_{ij}$ and using the matrices imposed by $\Delta(27)$ invariance:
\begin{equation}
I_{3s1} \equiv \mathrm{Im} \Tr (Y_{00} Y_{01}^\dagger Y_{02} Y_{00}^\dagger Y_{01} Y_{02}^\dagger) = 0\,.
\label{I3s}
\end{equation}
The CP symmetries present in these 12 special cases are discussed in \cite{Varzielas:2015fxa}.

Any choice of 4 or more singlets includes combinations of 3 that allow CP violation. By adding any other singlet to the set $h_{00}$, $h_{01}$, $h_{02}$ in $\mathcal{L}_{3s_1}$, we have a singlet $h_{ij}$ with $i \neq 0$. In general there is no vanishing of the $I_{3s}$-type CPIs involving $Y_{ij}$ with $Y_{00}$, $Y_{01}$, $Y_{02}$:
\begin{align}
\mathrm{Im} \Tr (Y_{00} Y_{01}^\dagger Y_{ij} Y_{00}^\dagger Y_{01} Y_{ij}^\dagger) \,, \\
\mathrm{Im} \Tr (Y_{01} Y_{02}^\dagger Y_{ij} Y_{01}^\dagger Y_{02} Y_{ij}^\dagger) \,, \\
\mathrm{Im} \Tr (Y_{02} Y_{00}^\dagger Y_{ij} Y_{02}^\dagger Y_{00} Y_{ij}^\dagger) \,.
\end{align}

\subsubsection{$\Delta(27)$ three triplets and beyond}

The next step is to investigate Yukawa-like Lagrangians in the presence of 3 $\Delta(27)$ triplets.
Considering Higgs doublets $h_u \sim 1_{10}$ and $h_d \sim 1_{01}$ (the notation is slightly different from the notation in other sections), and a $Z_2$ symmetry that ensures $u^c$ couple only to $h_u$ and $d^c$ couple only to $h_d$, the Lagrangian is:
\begin{equation}
\mathcal{L}_{2HDM} =y_u (Q u^c)_{20} h_{u} + y_d (Q d^c)_{02} h_d + H.c. \,,
\label{L2HDM}
\end{equation}
where $Q$ is a triplet and $d^c$, $u^c$ are anti-triplets of $\Delta(27)$.
Again I take Yukawa matrices $Y_u$, $Y_d$ corresponding to the terms and apply the IA.
The general CP transformations are denoted as
\begin{align}
h_u \rightarrow e^{i p_u} h_u^* \,, \quad h_d \rightarrow e^{i p_d} h_d^* \,, \\
Q \rightarrow U_Q^T Q^* \,, \quad u^c \rightarrow U_u u^{c *} \,, \quad d^c \rightarrow U_d d^{c *}\,.
\label{2HDMCP}
\end{align}
CP invariance demands, similarly to the SM case \cite{Bernabeu:1986fc}:
\begin{align}
U_Q Y_u U_u e^{i p_u}= Y_u^* \,,\\
U_Q Y_d U_d e^{i p_d}= Y_d^* \,.
\label{YuYd2HDM}
\end{align}
so I use the Hermitian combinations $H_{u,d} \equiv Y_{u,d} Y_{u,d}^\dagger$
\begin{align}
U_Q H_u U_Q^\dagger= H_u^* \,, \quad U_Q H_d U_Q^\dagger= H_d^* \,,
\label{HuHd2HDM}
\end{align}
concluding
$\Tr \left[ H_u, H_d \right]^3 = 0$   \cite{Bernabeu:1986fc} is necessary and sufficient for CP conservation. As $\Delta(27)$ imposes
\begin{equation}
Y_u=y_u
\begin{pmatrix}
1 & 0 & 0\\
0 & \omega^2 & 0\\
0 & 0 & \omega
\end{pmatrix} \,,
\end{equation}
\begin{equation}
Y_d=y_d
\begin{pmatrix}
0 & 1 & 0\\
0 & 0 & 1\\
1 & 0 & 0
\end{pmatrix} \,,
\end{equation}
CP is automatically conserved for any $y_u$, $y_d$.
In order to enable CP violation through an $I_{3s}$-type CPIs not vanishing requires at least 3 singlets coupling to one of the sectors, meaning that it is now possibly to have up to 6 distinct singlets and automatically conserve CP.

Beyond 3 triplets, more and more singlets can be included while the Lagrangian automatically conserves CP.
The final generalisation I consider is to add another anti-triplet $x^c$. If the sectors are separated by an Abelian symmetry (like the $Z_2$ discussed for $\mathcal{L}_{2HDM}$), there are 3 sectors of $\Delta(27)$ singlets that I denote $h_{d_{ij}}$, $h_{u_{kl}}$, $h_{x_{mn}}$. Using the IA and considering how CPIs can be constructed we extend the previous results to conclude that the relevant CPIs are of $I_{3s}$-type for each sector (due to the different $U_d$, $U_u$, $U_x$ matrices):
\begin{equation}
\mathcal{L}_{4Q} =Y_{d_{ij}} (Q d^c) h_{d_{ij}} + Y_{u_{kl}} (Q u^c) h_{u_{kl}} + Y_{x_{mn}} (Q x^c) h_{x_{mn}}  +  H.c. \,,
\label{L_4Q}
\end{equation}
\begin{align}
Q \rightarrow U_Q^T Q^* \,, \quad d^c \rightarrow U_d d^{c *} \,, \quad u^c \rightarrow U_u u^{c *} \,, \quad x^c \rightarrow U_x x^{c *} \,.
\label{4Q}
\end{align}

It is interesting that at 4 triplets (in this case 1 triplet and 3 anti-triplets) we have reached a situation where CP can be automatically conserved even with fields transforming as each of the 9 $\Delta(27)$ singlets. One example is $h_{d_{00}}$, $h_{d_{01}}$, $h_{d_{02}}$, $h_{u_{10}}$, $h_{u_{11}}$, $h_{u_{12}}$, $h_{x_{20}}$, $h_{x_{21}}$, $h_{x_{22}}$ \cite{Branco:2015gna}.

\section{Conclusions}

The invariant approach is very useful with family symmetries, and the examples I described serve to demonstrate this. One of the advantages of the method is that it does not depend on the group when the CP-odd invariants are constructed.
$\Delta(27)$ as a family symmetry has rich interplay with CP, which was also revealed through the examples that were explored.

I showed several Lagrangians, from cases with only 1-dimensional representations of $\Delta(27)$ (singlets), to Yukawa-like terms involving $\Delta(27)$ triplet and anti-triplet, and progressing to three and more $\Delta(27)$ triplets.

The number and type of representations fundamentally affects the CP properties of the Lagrangian. For those with only singlets, the invariant approach reveals the relevant physical phases, which turn out to be relative phases of the complex couplings. For the two triplet case (with one sector), CP is automatically conserved for Yukawa-like terms involving any 2 $\Delta(27)$ singlets and for 12 special combinations out the total 84 combinations of 3 singlets (the other cases are examples of explicit geometrical CP violation). The same type of conclusion holds independently for each sector, and therefore with 3 sectors it is even possible to have all 9 $\Delta(27)$ singlets present while automatically conserving CP.

\section*{Acknowledgments}

This project is supported by the European Union's
Seventh Framework Programme for research, technological development and
demonstration under grant agreement no PIEF-GA-2012-327195 SIFT.
I thank the organisers of Planck 2015 for hosting a very interesting conference.

\end{document}